\documentclass[12pt,preprint]{aastex}
\usepackage{amsmath}
\shorttitle{MHD Model for Formation Process of CMEs}
\shortauthors{Shiota, et al.}
\begin{document}
% \doublespace

\title{MHD Modeling for Formation Process of Coronal Mass Ejections:
Interaction between Ejecting Flux Rope and Ambient Field}

\author{Daikou Shiota\altaffilmark{1}, 
Kanya Kusano\altaffilmark{2,3}, 
Takahiro Miyoshi\altaffilmark{4}, 
Kazunari Shibata\altaffilmark{5}, 
%Nobuaki Ohno\altaffilmark{1}
}
\altaffiltext{1}{Advanced Science Institute, 
RIKEN (Institute of Physics and Chemical Research), 
Wako, Saitama 351-0198 Japan;
shiota@riken.jp
}
\altaffiltext{2}{Solar-Terrestrial Environment Laboratory, 
Nagoya University, Furo-cho, Chikusa-ku, Nagoya, 464-8601, Japan} 
\altaffiltext{3}{%The Earth Simulator Center,
Japan Agency for Marine-Earth Science and Technology (JAMSTEC),
Kanazawa, Yokohama, Kanagawa 236-0001, Japan}
%; 
%shiotadk@jamstec.go.jp}
\altaffiltext{4}{Graduate School of Science, Hiroshima University, 
Kagamiyama, Higashi-Hiroshima, Hiroshima, 739-8526, Japan} 
\altaffiltext{5}{Kwasan and Hida Observatories, Kyoto University,
Yamashina, Kyoto, 607-8471, Japan}

\begin{abstract}
We performed magnetohydrodynamic simulation of a
formation process of coronal mass ejections (CMEs),
focusing on interaction (reconnection)
between an ejecting flux rope and its ambient field.
We examined three cases with different ambient fields:  
no ambient field, 
and cases with dipole field of two opposite directions 
which are parallel and anti-parallel to 
that of the flux rope surface. 
As a result, 
while the flux rope disappears in the anti-parallel case, 
in other cases the flux ropes can evolve to CMEs and 
show different amounts of rotation of the flux rope.
The results imply
that the interaction between 
an ejecting flux rope and its ambient field
is an important process for determining 
CME formation and CME orientation,
and also show that 
the amount and direction of magnetic flux 
within the flux rope and the ambient field
are key parameters for CME formation.
Especially, the interaction (reconnection) plays
a significant role to the rotation of the flux rope,
with a process similar to ``tilting instability"
in a spheromak-type experiment of laboratory plasma.
\end{abstract}

%\keywords{Sun: magnetic field --- Sun: corona --- solar wind}

\section{Introduction}

Coronal mass ejections (CMEs) are 
one of the most spectacular explosive phenomena
in which large amounts of mass and magnetic flux 
are ejected to the interplanetary space.
CMEs are believed to be consequences of 
 sudden release of magnetic energy, 
i.e., disruptions of coronal magnetic field,
as same as flares which are strong brightening observed in soft x-ray.
The energy released upward can be converted to kinetic energy 
as ejected flux ropes and/or CMEs,
while energy released downward can be 
converted to thermal energy or high energy particles.
Therefore 
larger flares show higher association rate with CMEs 
\citep{1992ARA&A..30..113K, 2005ApJ...622.1240K,2005JGRA..11012S05Y},  
but they do not necessarily accompany each other. 
Association of flares can be explained by   
energy release rate which depends on the condition of coronal plasma.
If the energy release is fast enough to heat chromospheric plasma, 
the heated plasma fills above coronal loops, 
which are accompanied with strong brightening in soft x-ray observed as flares.  
If the energy release is not fast enough to overcome radiative energy loss,
the strong brightening does not occur. 
In the same time, if ejecting structures successfully escape to interplanetary, 
the event can be understood as CMEs without flares. 
On the other hand, 
in some cases, 
if an ejection associated with a strong soft x-ray brightening,  
cannot escape to the interplanetary space,
the event should be observed as a flare without CME
\citep{2006ApJ...650L.143Y}.

The disruptions of coronal magnetic field 
are observed to occur commonly in a wide spatial range. 
{\it Yohkoh} satellite reveals that 
faint giant coronal arcades which are similar to flares
are formed associated with CMEs 
\citep{1993GeoRL..20.2785H, 1996JGR...10113497M}.
On the other hand, the similar evolutions 
where flares occurs with mass (plasmoid) ejections
are observed accompanied much smaller scale emerging flux  \citep{sakajiri04}.
Based on such observational evidence,
\cite{1999Ap&SS.264..129S} proposed that  
giant arcades, flares, mass ejections, and CMEs
are understood with an unified view so-called 'flares'.
The model is supported by the statical characteristic of flares.
Solar flares show a ``power-law" distribution 
between their energy scales and occurrence rate 
\citep{1971SoPh...16..152D,1974SoPh...39..155D,
1985SoPh..100..465D,1995PASJ...47..251S}.

In contrast to flares,  
CMEs show  log-normal distribution
between the energy scales and the event number 
\citep{2004astro.ph..1352A,2005ApJ...619..599Y,2006JGRA..11106107L}.
Statistical characteristics of CMEs and flares 
are  different 
in spite of their common origin. 
This difference in statistical distribution 
means the existence of 
some kinds of filter effects which prevent
small scale mass ejections to evolve CMEs selectively.
Candidates of such filter effects are interactions 
between  eruptions and their ambient magnetic field
whose spatial scale is larger than that of the eruption regions. 
Important points in formation process of CMEs can be 
not only the formation of ejecting structures (flux rope)
but also whether they overcome the obstacles.  

Trigger processes of such eruptions 
have been numerically 
investigated for a few decade
\citep{
1995ApJ...446..377F,
1999ApJ...510..485A,2000ApJ...545..524C,
2004ApJ...610..537K,2006PhRvL..96y5002K}. 
In ``breakout" model \citep{1999ApJ...510..485A},
the interaction between a sheared arcade and its ambient field
is considered to play a fundamental role for the triggering. 
In the model, 
the field configuration is assumed to be 
favorable to reconnection 
between the erupting sheared arcade and its ambient field.
During the eruption 
the reconnection can reduce the poloidal magnetic flux of 
the flux rope 
which is supplied by another reconnection 
occurred inside the sheared arcade. 
As the results of competition between 
the magnetic flux reduction and the supply,
an eruption is capable of failing to produce a CME.
In some other models
\citep{1995ApJ...446..377F,
2004ApJ...610..537K, 2006ApJ...645..742I, 2006PhRvL..96y5002K}
ideal MHD processes 
(loss of equilibrium or instability) of detached flux ropes 
are dominant in the trigger processes.
In such models, the direction of its ambient field can be
 independent of the trigger processes. 
If the direction of the ambient field is favorable to reconnection, 
the reconnection can reduce the magnetic flux as mentioned above.
In contrast, if the direction is parallel to the poloidal flux 
the ambient field can drag and stop 
the ejecting flux rope 
\citep{1996ApJ...472..372C, 2008JGRA..11303S05S}.
Hence the interactions between ejecting flux ropes and their ambient field
can be a key factor for formation of CMEs.

From the point of view of space weather science, 
when CMEs containing southward magnetic flux
impact the Earth magnetosphere,
they make large influences 
on the space environment near the Earth.
Hence, it is also important for space weather forecast 
to understand the whole evolution process of CMEs;
how  CMEs are formed
and how much southward magnetic flux 
CMEs contain when they reach to the orbit of the Earth.
Estimation of southward magnetic flux 
is a complicated task important for space weather,
and requires understanding of initial conditions and 
whole evolution processes of eruptions.
For example, in nonlinear evolution of kink instability 
of a strongly twisted flux rope,    
its untwisting (writhe) motion can rotate the flux rope 
\citep{2004A&A...413L..23K,2005ApJ...630L..97T}.
The ejecting twisted flux ropes could change their directions 
due to the writhe rotations since their onset or formation 
if the ropes become unstable in the mode.  

Rotations of ejecting flux ropes are reported in 
 CME observations. 
\citet{2008ApJ...675L..49Y} investigated 
the angles of magnetic clouds, 
halo CME axes derived from coronagraph images, and 
EUV post-eruptive arcade axes in CME source regions.
They found that 
there are good correlation of angles between the axis of EUV arcades 
and CMEs,
and that between axes of magnetic clouds 
and coronal neutral lines (Heliospheric current sheets),
while there is low correlation of angles between CMEs and magnetic clouds. 
The fact means that 
some fraction of ejected magnetic structures
are significantly rotated 
during the travel from the solar corona to 1 AU.
During the way, the ambient field should be of two type of structure; 
closed and open.
In the  outer corona or interplanetary, 
all magnetic field should be open due to dragging of the solar wind 
while most of magnetic field lower corona should be closed.
The rotation may result from the interactions 
with the both type of ambient field. 
Especially, \citet{2008ApJ...675L..49Y} 
reported that a few events show the deviation between 
the neutral line angles of post eruptive arcades (in lower corona)
and the axes of CMEs (in a few solar radii).
This means that the rotation during the evolution in lower corona, 
i.e., interaction with closed field, 
is not negligible.

In this study, 
we performed a three-dimensional 
MHD simulation of a twisted flux rope 
ejected from an active region 
surrounded by a global ambient magnetic field.  
According to the numerical results,
we discuss the condition for formation of CMEs and 
evolutions of CME magnetic field structures, 
focusing on the interactions between
ejecting flux ropes and their ambient closed field. 

In the next section, we describe 
the detailed methodology of the numerical simulation.
In section 3, the numerical results of different 
ambient field are shown. 
We discuss the quantitative relations 
suggested by the numerical results in \S 4. 
Finally, we summarize this paper in \S 5.

\section{Numerical Model}
\subsection{Numerical Scheme}

We performed three dimensional  MHD simulation  
which solves time variation of 8 physical quantities; 
density, gas pressure, velocity and magnetic field 
($\rho, p, {\bf v}$, and ${\bf B}$, respectively). 
In order to simplify the MHD equations, 
the 8 physical quantities are normalized with 
the following typical values in the corona:
$L_0=R_{\odot}=6.99\times10^{5}$ [km],
$B_0= 30 $ [Gauss], 
$\rho_0= m_H \times 10^9 = 1.67 \times 10^{-15}$ [g cm$^{-3}$],
$v_0=V_{0,A}=B_0/\sqrt{4 \pi \rho_0}= 2071 $ [km s$^{-1}$],
as these results, $\tau_0=\tau_A=350 $ [s],
$p_0 = \rho_0 V_{0,A}^2 
= 7.16 \times 10^{1} $ [erg cm$^{-3}$].
%$p_0 = \rho_0 V_{0,A}^2/\gamma 
%= 7.16 \times 10^{1} / \gamma$ [erg cm$^{-3}$].
The normalized MHD equations 
(equation of continuity, equation of motion, induction equation, 
and equation of energy with gravity)
are expressed as follows:
\begin{equation}
{\frac{\partial \rho}{\partial t}}+\nabla \cdot (\rho {\bf v})=0,
\label{coeq2} 
\end{equation}
\begin{equation}
{\frac{\partial \rho {\bf v}}{\partial t}}
+ \nabla \cdot (\rho {\bf v}{\bf v} - {\bf B} {\bf B} + p_T {\bf I} ) = \rho {\bf g},
\label{moeq2} 
\end{equation}
\begin{equation}
\frac {\partial {\bf B }}{\partial t } + 
\nabla \cdot \left( {\bf B }{\bf v}-{\bf v}{\bf B }
+ \psi {\bf I} \right) = - \nabla \times  \left( \eta {\bf J }\right),
\label{ineq2} 
\end{equation}
\begin{equation}
{\frac{\partial e}{\partial t}} 
+ \nabla \cdot \left[ (e + p_T ) {\bf v} - ({\bf v} \cdot {\bf B} ) {\bf B} \right]=0,
\label{eneq2}
\end{equation}
where
\begin{gather}
{\bf J } = \nabla \times {\bf B }, \\
e = \frac{ \rho {\bf v}^2 }{2} + \frac{p}{\gamma -1} 
+ \frac{{\bf B}^2}{2} + \frac{G_0}{r} \rho, \\
p_T = p +  \frac{{\bf B}^2}{2}, \\
{\bf g} = - G_0 {\bf r} / r^3, \\
G_0 =  \frac{G M_{\odot} }{R_{\odot} v_{0}^2},
\end{gather}
$G$ is gravitational constant, 
${\bf I}$ is the unit matrix in the Cartesian coordinate,
$\psi$ is additional variables described below,
and $M_{\odot}$ is mass of the Sun.
Divergence operator is solved by 
a finite volume method 
with the HLLD nonlinear Riemann solver 
\citep{2005JCoPh.208..315M}
which is combined with
the 3rd order Monotone Upstream-centered 
Schemes for Conservation Laws
(MUSCL) developed by \citet{1979JCoPh..32..101V} and 
the 2nd order Runge-Kutta time integration.
Specific heat ratio $\gamma $ is set to be 1.05
which means an additional heat source. 
Resistivity $\eta$ is set to be $0$  
which means reconnection is caused by the numerical diffusion.  

Generally, the results of time integration of equation (\ref{ineq2})
multi-dimensional MHD simulation is possible to
cause numerical $\nabla \cdot {\bf B }$ 
due to discretization of   numerical domain.
The equations (\ref{moeq2}), (\ref{ineq2}), and (\ref{eneq2})
are derived with Solenoidal condition, 
and therefore 
such finite numerical $\nabla \cdot {\bf B }$ can 
break down the simulation. 
In order to keep the numerical $\nabla \cdot {\bf B}$ to 
be sufficiently small value,
we applied an additional variable $\psi$ and 
additional equation, 
\begin{equation}
\frac {\partial \psi }{\partial t } + c_h^2 \nabla \cdot {\bf B } + c_h^2/c_p^2 \psi =0
\label{eqpsi} 
\end{equation}
based on the hyperbolic divergence B 
cleaning method \citep{2002JCoPh.175..645D}.
The coefficient $c_h$ in Equation (\ref{eqpsi}) is 
propagation speed of the numerical $ \nabla \cdot {\bf B}$,
and the coefficient $c_p$ is diffusion coefficient of $\psi$.
This numerical methods are described more in 
\citet{2008JGRA..11303S05S}.

The numerical domain is a spherical shell of 
$(r_{\rm min}, r_{\rm max}) = (1.01, 5.0)$.
The domain is discretized with
Yin-Yang grid %(Kageyama \& Sato 2004), 
\citep{2004GGG.....5.9005K},
a chimera grid composed of 
two congruence spherical partial shells 
$(\pi/4 \leq \theta \leq 3\pi/4 $,$ -3\pi/4 \leq \phi \leq 3 \pi/4)$  
each of which is installed in different direction.
The physical quantities on the edges of the each partial shell 
are substituted by interpolated values 
at the same position in the counterpart shell. 

Thus the boundary condition in this grid system 
only at inner and outer edges for the radial direction.
The inner boundary is assumed to be the line-tying wall,
where $B_{\rm norm}$ is constant and ${\bf v} = 0$. 
The outer boundary is assumed to be the free boundary,
where the radial gradients of all quantities are zero.

Note that, in this simulation, test particles 
which are distributed in some manners in initial condition 
are advected with plasma bulk flows.  
The magnetic field lines in Figures in this paper
are obtained with the integration of the magnetic field 
from the position of the test particles of each time step.
Thus we capture the evolution of individual magnetic field lines 
with this technique.

\subsection{Initial condition}

The subject of the present work is to understand
how a flux rope ejected from an active region 
evolves to a CME.
The initial magnetic field consists of 
the ejecting flux rope (${\bf B}_{\rm S}$)
and ambient global field (${\bf B}_{\rm D}$).%;
We performed three cases of simulation 
where only the ambient field $ {\bf B}_{\rm D}$ is different.
The ambient global field (${\bf B}_{\rm D}$)
is chosen to be simple dipole field: 
\begin{align}
B_{r,{\rm D}}(r,\theta,\phi) =&  \frac{2  B_{\rm D} \sin \theta}{3 r^3} \\
B_{\theta,{\rm D}}(r,\theta,\phi) =& \frac{ B_{\rm D} \cos \theta }{3 r^3}.
\end{align}
In order to study the interaction 
between an ejected flux rope and ambient field, 
we carried out three cases of simulation 
with different strength and direction of the ambient fields, 
i.e., their amplitude $B_{\rm D}$ of 
$   0.0 $,$ - B_{\rm D,0}$,  and $ B_{\rm D,0}$, 
which are labeled with case A, B, and C, respectively.
$B_{\rm D,0}=0.124 $ (3.72 [G])
is a base field strength of the corona
where the Alfv{\'e}n speed is equal to the sound speed of $T = 2 \times 10^6 $ [K].

\begin{figure}
\plotone{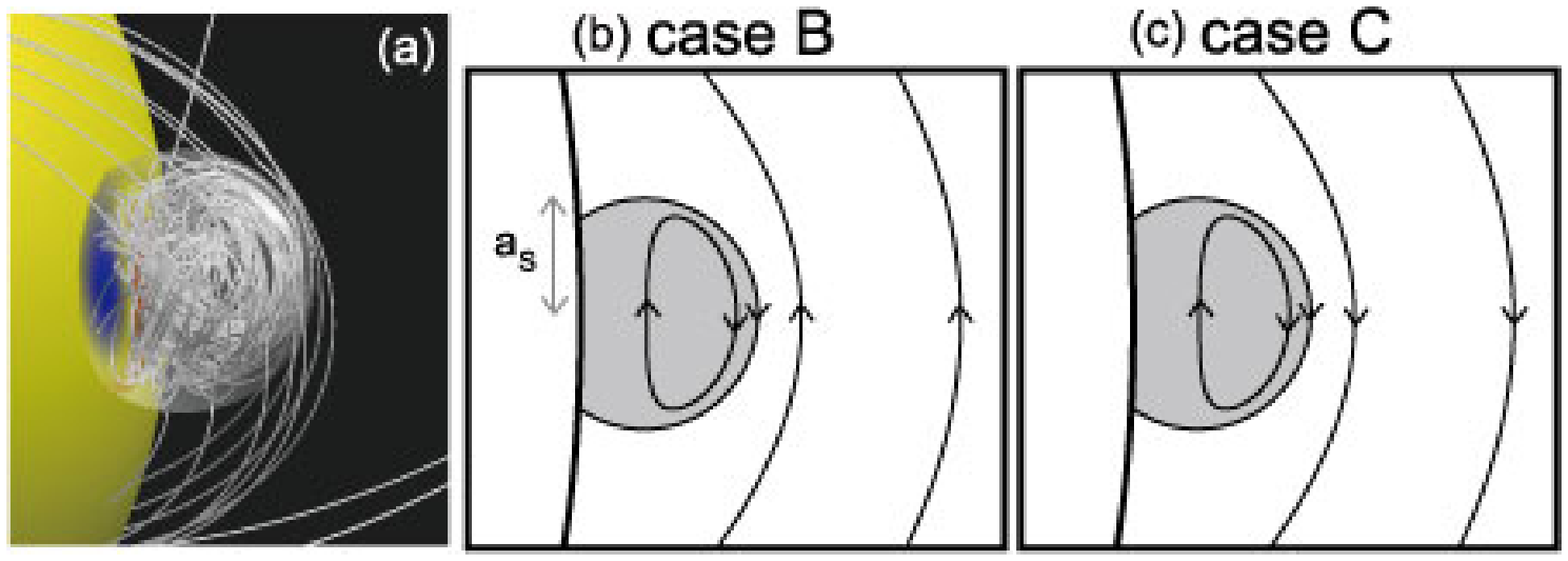}
\caption{
(a) Three dimensional view of initial magnetic field
in case C. 
The surface color shows $B_r$ map and 
tubes are magnetic field lines. 
Transparent surface illustrates 
the surface boundary of the spheromak field.  
(b) Schematic picture of initial 
magnetic configuration in the $y=0$ plane in case B.
Spheromak type field of whose radius is $a_S$ are assumed 
in the initial condition (shadowed region).
The solid lines express magnetic field lines projected on the $y=0$ plane. 
(c) Schematic picture of initial magnetic configuration in case C 
shown with the same way as panel (b).
}
\label{initmf}
\end{figure}

We mimic the ejected flux rope formed 
as a result of eruption in an active region 
with 
a spheromak type magnetic field which is 
a linear force-free field in a completely 
isolated sphere (see Figure \ref{initmf}a).
\begin{equation}
{\bf B }_{\rm S}(r,\theta,\phi) 
= \tilde{{\bf B}}(\tilde{r},\tilde{\theta},\tilde{\phi}),
\end{equation}
where $\tilde{{\bf r}} = \alpha ({\bf r} - {\bf r}_{\rm S}) $ is
the translated and rescaled local spherical coordinate
whose origin corresponds to the center of the spheromak. 
The spheromak type field is expressed as follows: 
\begin{align}
\tilde{B}_{\tilde{r}}(\tilde{r},\tilde{\theta},\tilde{\phi})=& 
- 2 B_{\rm S0}
\frac{ j_1 \left( \tilde{r} \right) }{ \tilde{r} }
\cos \tilde{\theta}, \label{eqbr}\\
\tilde{B}_{\tilde{\theta}}(\tilde{r},\tilde{\theta},\tilde{\phi})=& 
 B_{\rm S0} \left(
\frac{ j_1 \left( \tilde{r} \right) }{\tilde{r} }
+ j_1^{\prime} \left( \tilde{r} \right)
\right) \sin \tilde{\theta}, \label{eqbt}\\
\tilde{B}_{\tilde{\phi}}(\tilde{r},\tilde{\theta},\tilde{\phi})=& 
B_{\rm S0}
 j_1 \left( \tilde{r} \right) 
\sin \tilde{\theta}, \label{eqbp}
\end{align}
where $B_{S0}$ is field strength of the spheromak 
and $j_1$ and$j_1^{\prime}$ are first order spherical
Bessel function and its derivative:  
\begin{align}
j_1 \left( x \right) = & \frac{\sin x - x \cos x }{x^2}, \\
j_1^{\prime} \left( x \right) = & 
\frac{2 x\cos x - \left(x^2 -2 \right) \sin x }{x^3}.
\end{align}
We adopt only the field within the isolated spherical shell 
$ \tilde{r} \leq \tilde{a} $ 
(transparent sphere in Figure \ref{initmf}a)
while  $\tilde{\bf B}_{\rm S} =0 $ outside of the shell.
As definition of the spherical Bessel function,
$  \tilde{a} = 4.493409458 $ and  
$ \alpha  = \tilde{a} / a_{\rm S}  $. 
The toroidal flux ($\Phi_{\rm S}$) and 
poloidal flux ($\Phi_{\rm P}$) of this field 
are obtained by numerical integrations;
\begin{gather}
 \Phi_{\rm S} =  \int_0^\pi {\rm d}\tilde{\theta} 
\int_0^{a_{\rm S}} \tilde{r} \tilde{B}_{\tilde{\phi}} {\rm d}\tilde{r} 
 = 0.261 a_{\rm S}^2 B_{\rm S0},
 \label{ftor}\\
 \Phi_{\rm P} =  
\int_{-\pi}^{\pi} {\rm d}\tilde{\phi} \int_0^{a_{\rm S}} 
\frac{\tilde{r}}{2}
\left|\tilde{B}_{\theta}
\left(\tilde{\theta} = \frac{\pi}{2}\right)
\right| {\rm d}\tilde{r} 
 = {0.0736 a_{\rm S}^2} B_{\rm S0}.
\label{fpol}
\end{gather}

In the present paper, we choose the position of the center of the spheromak
 ${\bf r}_{\rm S} = (1.1,\pi/2,0)$ in the spherical coordinate
(on the $x$ axis), and
the major axis of turns is parallel to the $z$ axis. 
The radius of the spheromak $ a_{\rm S} = 0.2 $, 
i.e., $ 1.4 \times 10^{10}$ [cm]. %so that
Here, we assume the toroidal flux of the spheromak 
$ \Phi_{\rm S} = 0.0174 $ 
$ = 2.55 \times 10^{21} $ [mx]
which is a typical amount of magnetic flux within a CME
\citep{2009JGRA..11410102K}.  
Then Equation (\ref{ftor}) yields
the field strength of the spheromak 
$ B_{\rm S0} =  \Phi_{\rm S}/(0.261 a_{\rm S}^2)= 1.67 = 50 $ [Gauss].  

Although a spheromak is a linear force-free field, 
we adopt it only within the boundary $\tilde{r} \leq \tilde{a} $.
At the spherical boundary $\tilde{r} = \tilde{a} $,
the local radial ($\tilde{B}_{\tilde{r}}$) and 
azimuthal ($\tilde{B}_{\tilde{\phi}}$) components vanish
and then only the local zenithal component $\tilde{B}_{\tilde{\theta}}$
exists
while $\tilde{\bf B} =0 $ outside of the boundary.
Therefore the force does not balance at the spherical boundary,
i.e., a strong outward gradient of magnetic pressure exists there.
The  Lorentz force due to the current
which corresponds to a steep magnetic pressure gradient 
causes the strong expansion.
The spheromak field naturally swells in the simulation.

Figures \ref{initmf} shows schematic pictures of 
magnetic field direction
on the meridional plane through the center of 
in cases B (\ref{initmf}b) and C (\ref{initmf}c).
The spheromak local zenithal field 
on its surface $ \tilde{r} = \tilde{a} $ satisfies  
$\tilde{B}_{\tilde{\theta}}  \left( \tilde{r} = \tilde{a} \right) 
\geq 0 $ (cf. eq. (\ref{eqbt})),
which is southward.
On the other hand, the ambient field is  
northward ($B_{\rm D} < 0 $) in case B, 
while it is southward in cases C.
On the equator, 
$\tilde{B}_{\tilde{\theta}}\left( \tilde{r} = \tilde{a} \right)$ 
and $B_{\theta,{\rm D}} < 0 $ becomes anti-parallel in case B,   
while parallel in case C.
Hence, cases A, B, and C are referred as `no ambient' 
`anti-parallel', and `parallel' cases in this paper.

Density and pressure are determined by 
hydrostatic equilibrium with uniform temperature 
$T = 2 \times 10^6 $ [K], 
\begin{gather} 
 p = p_{\rm b} \exp\left[ \frac{G_0 \gamma}{p_{\rm b}}   
\left( \frac{1}{r} - 1   \right)
\right] \\
\rho = \gamma \frac{p}{p_{\rm b}} \\
{\bf v} =0, 
\end{gather}
where $p_{\rm b} = 2  \rho_0  k_{\rm B} T /  ( m_H  p_0 )$ 
is the normalized pressure on solar surface ($r=1$). 
In the present study, we did not apply the solar wind,
which is also important factor for the formation of a CME. 
This is because we intend to purely investigate 
the interaction between 
the ejecting and the ambient fields
in the early phase of the CME formation. 

\section{Numerical Results}
\label{res}

\subsection{Common evolution}
\label{commn}
In all cases,
the spheromak (flux rope) initially swells outward
due to the non-equilibrium initial condition
and the line-tying boundary
as described in the previous section. 
At the same time,
the expanding flux rope also shows 
the similar rotation around 
the line of the ejection in all case.
This rotation is caused by an untwisting (writhe) motion of
the highly twisted initial spheromak structure 
(see Figure \ref{initmf}a).

In order to show the height-time evolution, 
the right panels of Figure \ref{albta} illustrate 
time variation of plasma 
$\beta \equiv 0.5 p/ B^2 $ 
along the $x$ axis. 
% In the present simulation, 
% because specific heat ratio $\gamma $ is assumed to be 1.05, 
% the decrease in pressure 
% due to the expansion of the flux rope is mild.
% Therefore, the plasma $\beta$ inside 
% the flux rope slightly increases.
The low plasma $\beta$ regions along the $x$ axis
express the positions of the flux rope.  
Therefore the low plasma $\beta$ in the height-time diagrams 
illustrated in Figure \ref{albta}  
corresponds to the trajectories of the flux ropes.
In all the cases
the flux rope rapidly expands just after the start,
followed by  
rising and expanding motion with some deceleration.

In all the cases,
many small features appear around the altitude $r \sim 1.1$
and rise as time below the flux rope 
in the right panels of Figure \ref{albta}. 
These structures are small plasmoids formed by tearing of 
the current sheet between the anchored axial flux.
Once such plasmoids are formed in the current sheet,
due to gradient of magnetic pressure,
they are ejected to upper direction, and 
finally collide and coalesce into the above flux rope. 
However, the momentum and energy additions by the plasmoids 
are too small to accelerate the flux rope.
The evolution is similar to the results of 
2D simulation \citep{2008JGRA..11303S05S}.

\subsection{`No ambient' case}
\label{noamb}
In `no ambient' case (case A), 
the flux rope can expand into unmagnetized plasma.  
We simulated this case as a reference case 
with no magnetic obstacle.
The expanding flux rope drags outward its anchored axial flux. 
The dragged axial flux forms $\Omega$-shape and 
at the center of them a current sheet is formed.   
Although the reconnection occurs 
between the dragged fluxes, 
a part of the flux remains anchored.

Figures \ref{albta}a and \ref{albta}b show 
the magnetic field structure
at $t=50$ (well-developed phase) and 
the height-time diagrams in this case.
Figure  \ref{albta}b expresses 
the time evolution of the flux rope in this case.
The flux rope  rapidly expands just after the start 
($t \lesssim 10$: early phase),
and then it shows  
rising and expanding motion during 
middle phase ($10 \lesssim t \lesssim 30$).
The flux rope is slightly decelerated.
In the late phase ($t \gtrsim 30 $), 
the flux rope  
expands self-similarly at almost a constant speed,
and finally reaches  
the outer boundary of the numerical domain.

\subsection{`Anti-parallel' case}
\label{anti}
We simulated the `anti-parallel' case (case B) as 
a condition where the directions of poloidal fluxes 
of the ejection and the ambient field are opposite. 
Such a condition is probable 
if the flux rope is formed and ejected in 
a newly emerged flux system whose direction is
opposite to the pre-existing ambient field.
This situation is modeled 
as a candidate system for 
rapid eruption for a CME
\citep{1999ApJ...510..485A}.

The time evolution of the flux rope 
is significantly different from that in case A.
Because the direction of the ambient field is 
opposite to the poloidal field of the ejecting flux rope,
there is a strong current sheet 
where magnetic reconnection occurs. 
Similar to the scenario of \citet{1999ApJ...510..485A},
the rise motion of the flux rope 
presses the current sheet, 
and then enhances its current density, 
i.e., the reconnection rate there.
However, 
the poloidal flux of the flux rope is 
peeling off due to the reconnection with the ambient
as it rises.
%It is thought that
Finally, as shown in Figure \ref{albta}c,
after most of the poloidal flux is reconnected,  
the flux rope no longer exists. 
The magnetic flux becomes just coronal loops 
within which shear Alfv{\'e}n waves propagate.
Such a result may be a candidate condition for
plasmoid ejections (flares) without CMEs. 

We can see the time evolution of the flux rope
in height-time diagram (Figure  \ref{albta}d). 
In the early phase ($t \lesssim 10 $), 
the trajectory and the vertical size are similar to those in case A.
After the phase, the top edge of the flux rope
becomes lower than that in case A because of the flux peeling
($10 \lesssim t \lesssim 35 $: middle phase). 
Finally ($t \gtrsim 35$: late phase), 
the plasma $\beta$ inside the flux rope 
becomes comparable to unity, it stays almost the same height.  
In this phase, 
the structure is recognized as 
twisted loops rather than a flux rope.

\subsection{`Parallel' case}
\label{para}
We simulated the `parallel' case (case C) as 
a condition where the directions of poloidal fluxes 
of the ejection and the ambient field are parallel. 
CME formation process in a similar but axisymmetric situation 
is studied by 
\cite{2008JGRA..11303S05S}.
The present study is an extension of \cite{2008JGRA..11303S05S} 
to study three-dimensionality of CME formation process.

\begin{figure}
\plotone{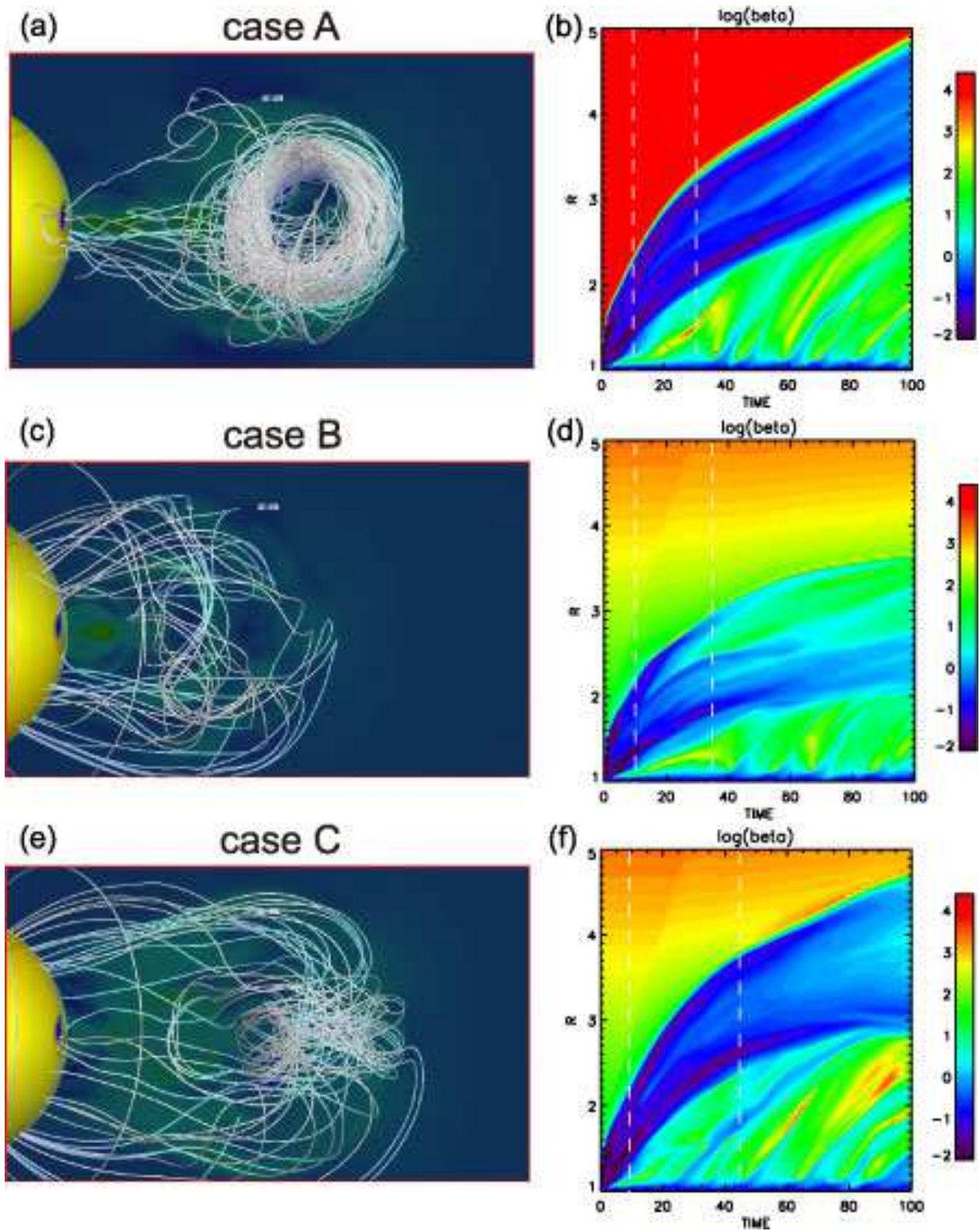}
\caption{(a) Field configurations in late stages ($t= 50$) in case A (no ambient). 
(b) timeslice of plasma $\beta$ on the $x$ axis in case A (no ambient). 
(c) and (d) are those in case B (anti-parallel), 
and (e) and (f) are those in case C (parallel), respectively.
In the timeslice images, the trajectories of the flux rope 
are recognized as low $\beta$ region with rising motion and
white dashed lines indicate timing 
when the evolution of the flux rope changes describes in \S 3. }
\label{albta}
\end{figure}

The time evolution of the flux rope 
is a little different from that in case A,
as shown in Figure \ref{albta}e and \ref{albta}f. 
In the early phase ($t \lesssim 10 $), 
the flux rope rapidly expands, and  
shows rising and expanding motion during 
in the middle phase ($10 \lesssim t \lesssim 45$),
and then a self-similar evolution in the late phase ($t \gtrsim 35$).
The expanding flux rope forms an $\Omega$-shape loop and 
its central current sheet where magnetic reconnection occurs,
and finally the flux rope reaches the outer boundary,
similar to in case A.  
In contrast to the result of case A,
the ejecting flux rope is 
significantly deformed 
(Figures \ref{albta}a and \ref{albta}e),
although the flux rope is 
successfully ejected to the outer boundary in both cases.
The difference should be caused by reconnection between 
the flux rope and the ambient magnetic field. 
%The  height-time diagram is shown in Figure  \ref{albta}f.
The height evolution of the flux rope  
is similar to that in case A
but the flux rope continues to be decelerated
(Figures \ref{albta}b and \ref{albta}f).

\begin{figure}
\plotone{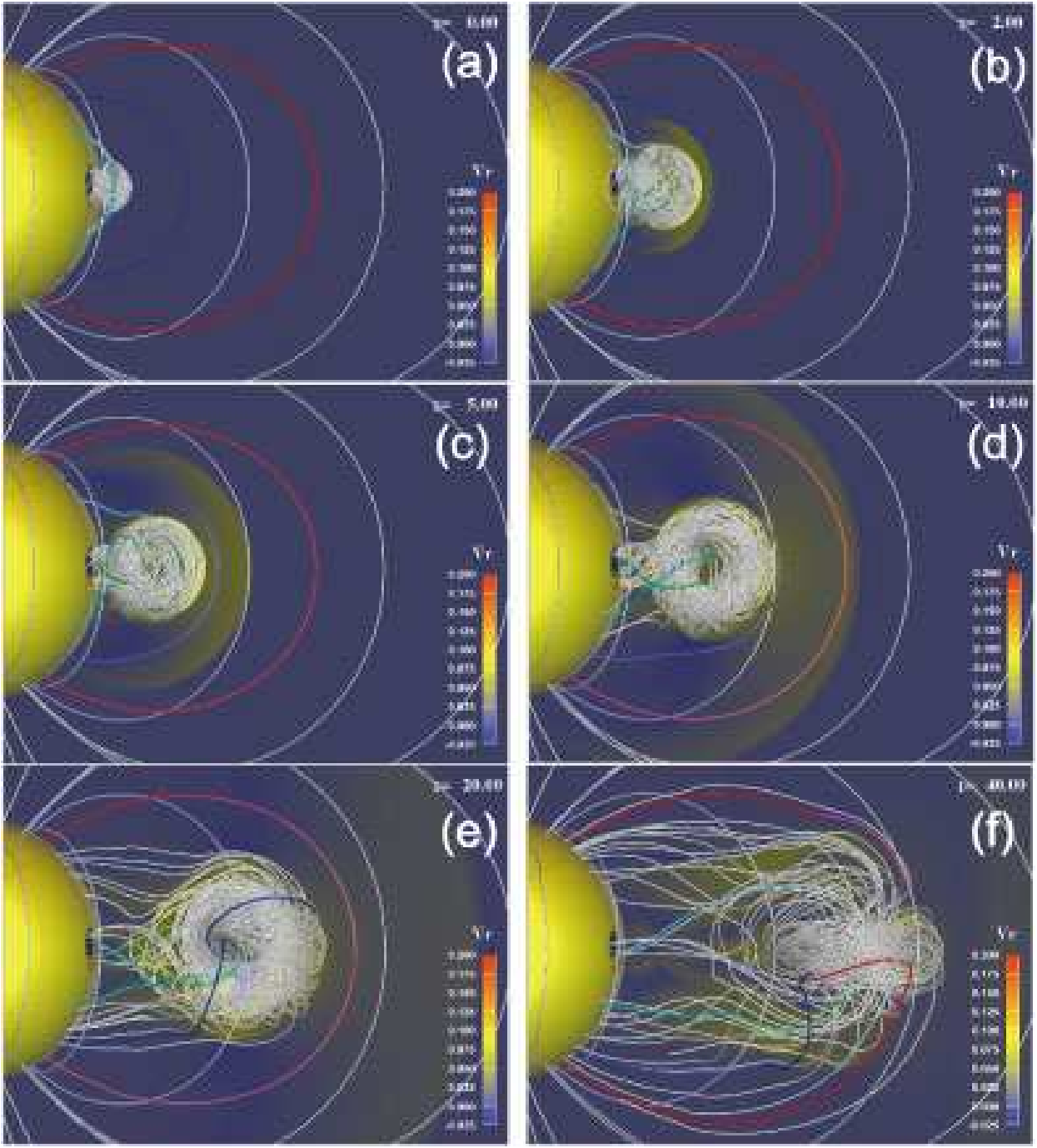}
\caption{Time evolution of magnetic field and velocity in case C.
Panels (a) - (f) show those at $t = $0, 2, 5, 10, 20, and 40, 
respectively. 
The inner boundary is shown with the sphere 
in the left part of each panel,
whose colors indicate $B_r$ map.
The same field lines are displayed with tubes in each panel.
Characteristic field lines are colored with
green, light blue, blue, and red. 
The velocity map at $ y = 0 $ are displayed with colors 
on the vertical transparent background. 
}
\label{evoC}
\end{figure}

\begin{figure}
\epsscale{0.75}
\plotone{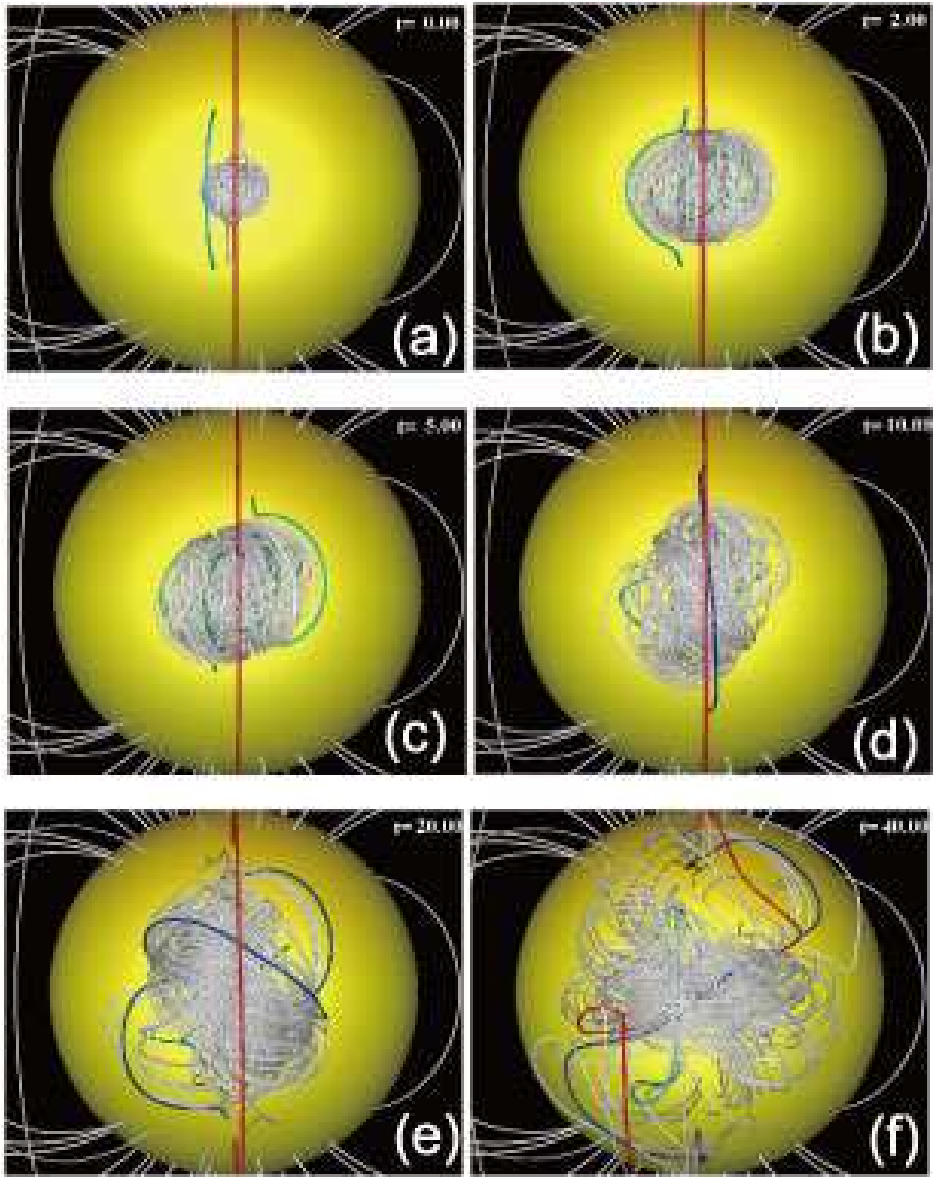}
\caption{Time evolution of face-on view of magnetic field structure in case C.
Panels (a) - (f) show those at $t = $0, 2, 5, 10, 20, and 40, respectively.
The meaning of the colors on the sphere and the tubes are 
the same as those in Figure \ref{evoC} }
\label{angle}
\end{figure}

We investigated the time evolution of 
the magnetic field configuration.
Figure \ref{evoC} and \ref{angle} shows the 
time evolution of the magnetic field configuration
seen from the side and the face of the ejecting flux rope.
Some characteristic field lines are 
highlighted with colors (green, light blue, blue, and red) 
in the figures.
These lines are drawn by integrating magnetic field 
from the same points on the inner boundary 
in each time step.

The ``green" field line represents 
the initial axial field line.
As the flux rope expands, 
the axial field lines are dragged and stretched 
as shown in Figure \ref{evoC}a - \ref{evoC}e. 
Finally the field line reconnects 
at the $\Omega$-shaped center as described in \S \ref{noamb}.

The ``light blue" field line indicates ``interchanged field line". 
The line is initially an ambient field line 
(Figure \ref{evoC}a and \ref{angle}a). 
As the flux rope expands,
the line is pushed aside by the flux rope 
(Figure \ref{evoC}b and \ref{angle}b).
Because the direction of magnetic field is parallel,
the ambient field line does not reconnect with the flux rope. 
However, as the flux rope evolves, 
it rotates due to writhe motion. 
The field lines are connected to inside of the flux rope
(Figure \ref{evoC}c and \ref{angle}c).
The result implies that the field lines reconnect with the flux rope
at the current sheets formed on the flux rope surface. 
The location of the 
reconnection is examined later (\S \ref{dis2}).

The ``blue" field line indicates 
also ``interchanged field line". 
The line is also initially an ambient field line
further than ``light blue" line.
The ``blue" field line reconnects 
with the ejecting flux rope 
at the current sheet formed on its surface 
as same as the ``light blue" field line.

Reconnection occurs also in the front of the flux rope.
In the initial condition,
the poloidal field of the surface of the flux rope 
is parallel to the ambient field.
As mentioned in \S \ref{commn}, however, 
the flux rope rotates  
due to its writhe motion. 
The direction of magnetic field on the flux rope front 
rotates counterclockwise as shown in Figure \ref{angle}.
The rotation cause the formation of a current sheet on the front
and then magnetic reconnection occurs there.

See the ``red" field line in Figure \ref{evoC} and \ref{angle}.
The ``red" line was initially ambient dipole field
but finally  connected to
the inside of the flux rope, i.e.,
becomes the new axial field of the ejecting flux rope
(Figure \ref{evoC}f and \ref{angle}f).

\section{Discussion}

\subsection{Magnetic Flux relation for CME formation}
\label{flux}
The amounts of magnetic flux and magnetic helicity are 
key parameters to understand the numerical results.
Here we estimate the amounts of magnetic flux and helicity 
and discuss the relation between them and 
a condition for CME formation.
In the initial conditions of case B and C, 
though the spheromak field structure is superposed into the ambient dipole field, 
the ambient dipole field component is so weak that it does not change the spheromak field.
Hence we can regard the time evolutions in case B and C as 
the results from the interaction (reconnection) 
between ejecting flux rope and the ambient field.

Whether an ejecting flux rope can evolve to a CME depends on 
the magnetic flux relation of the overlying ambient field and 
the ejecting flux rope.
In the present study, the dipole field is assumed 
as the ambient field
and therefore we can integrate the amount of the flux 
with the following equation:
\begin{equation}
\Phi_{\rm D} =\int_{\phi_{\rm min}}^{\phi_{\rm max}} {\rm d}\phi 
\int_{r_{\rm min}}^{r_{\rm max}} B_{\theta,{\rm D}}
\left(\theta = \frac{\pi}{2}\right) r {\rm d}r. 
\end{equation}
Because only the magnetic flux above 
the spheromak field is relevant to the CME condition, 
the azimuthal width $\phi_{\rm min},\phi_{\rm max}$ 
of the overlying magnetic flux
are determined by that of the spheromak field 
in the initial condition:
$\phi_{\rm min},\phi_{\rm max} = 
\pm \tan^{-1} (a_{\rm S} /|{\bf r}_{\rm S}|)$.
Then we obtain the overlying magnetic flux 
$\Phi_{\rm D} = 1.19 \times 10^{-2} $.
On the other hand we can also integrate the poloidal flux 
of the spheromak field with the equation (\ref{fpol}).
The azimuthal width in spheromak coordinate is set to $\pm \pi/2 $ 
when the upper half of the spheromak field is assumed to
commit to their ejection.
Then we get $\Phi_{\rm P} = 2.45 \times 10^{-3} $. 
The relation between amount of magnetic flux  
$ \Phi_{\rm P} < \Phi_{\rm D} $ in case B and C.
This condition naturally explains the  
disappearing of the flux rope 
due to flux peeling-off reconnection in case B.
In case C,
the reconnection is not completely anti-parallel 
and so inefficient that 
the amount of the stripped magnetic flux is 
less than that of the flux rope.
In case A, there is no obstacle ambient field. 
Hence, the flux rope in case A and C is not dissipated.

Although only three typical cases are modeled in the present study, 
there can be any intermediate situations.
For example, the ambient field is anti-parallel 
but total amount of magnetic flux is smaller than that of an ejecting flux rope,
the flux rope also retains its shape and escapes to the outer space.
Such a condition is a candidate condition where explosive energy release
has been studied as the breakout model \citep{1999ApJ...510..485A}.
In the case of small scale, 
\cite{2007ApJ...666..516G} and \cite{2008A&A...492L..35A}
studied emerging flux into preexisting and weak coronal field.
\cite{2007ApJ...666..516G} showed rising motion of the emerging flux
is not affected by the orientation between the emerging flux 
and the preexisting field.  
\cite{2008A&A...492L..35A} showed in anti-parallel case 
the emerging flux rope successfully erupts 
while remains confined in other case. 
In such cases of emerging flux, 
the amount of emerged magnetic flux becomes 
more than that of the coronal field within the same spatial scale,  
because the strength of the magnetic field is 
very strong in the lower atmosphere. 
However,  
since the magnetic field in the real Sun is very complicated, 
magnetic field in small scale does not necessarily have
same direction with global scale field. 
Furthermore the spatial scale of emerging flux 
should be (sometimes very) small compared to the solar radius,
the flux of global scale ambient field 
is possibly comparable or larger than that of flux ropes.

In this study, we assume a spheromak field structure 
smaller than the global scale (the solar radius) and
which has a typical amount of magnetic flux 
$ 2.553 \times 10^{21} $ [mx].
The ambient field $\sim 2.5 $ [G] is also 
comparable to or
a little weaker than a typical coronal value. 
The difference in the direction and the amount  
of the overlying field leads to 
much different results.
This result may give a hint to understand the observational result that 
all X-class flares are not necessarily associated with CMEs
\citep{2005JGRA..11012S05Y}.

Hence, the overlying field above an eruption region
is an important factor for CME formation, 
and therefore its detail estimation 
with a realistic corona model is required for 
CME prediction for space weather forecast.
Note that we also did not consider the effect of solar wind
which makes substantial amount of magnetic flux  open 
and decreases magnetic obstacle.
The detail estimation of amount of magnetic flux
using observational data and a realistic corona model is 
an important task for a CME prediction.

\cite{2008JGRA..11303S05S} 
discussed a condition for CME formation with 2D simulation of 
continuously sheared arcade which results in
a CME or a confined flux rope formation.
They 
suggested that the condition for CME formation depends on
the balance between amounts of 
the magnetic helicity ($H$) within the flux rope and sheared arcade, 
and overlying magnetic flux ($\Phi_{\rm ov}$)
which confines the sheared structure, as follows;
\begin{equation}
  \frac{\Phi_{\rm ov}^2}{|H|} < 1.7.
\label{shio08}
\end{equation}
In these symmetric condition, 
maximum energy for the flux rope to escape 
requires very much due to the work to drag the overlying flux. 
Therefore the condition of equation (\ref{shio08}) is 
possibly a upper limit for a 3D condition. 

The relative magnetic helicity of the each case of the present study,
is estimated from the initial condition 
\begin{equation}
H =  \int \left({\bf A }+{\bf A}_{\rm p} \right) 
    \cdot \left( {\bf B} -{\bf B}_{\rm p} \right){\rm d}V,
\end{equation}
where ${\bf B}_{\rm p}$ and ${\bf A}_{\rm p}$ are 
the potential magnetic field 
and the corresponding vector potential 
which are determined
by the magnetic field radial component 
on the inner boundary of the numerical domain.
The estimated relative helicity is
$H=-3.65\times10^{-4},-2.96\times10^{-4},-4.34\times10^{-4}$ 
in case A, B, and C, respectively.
The relation of equation (\ref{shio08}) for case becomes 
\begin{equation}
 \frac{\Phi_{\rm ov}^2}{|H|} =  \frac{\Phi_{\rm D}^2}{|H|} 
= \frac{(1.19 \times 10^{-2})^2}
{4.34\times10^{-4}} \sim 0.33 < 1.7.
\end{equation}
This relation suggests that 
the confinement due to the overlying field
is inefficient to prevent a flux rope escape in case C 
the results are consistent with the results of 
\cite{2008JGRA..11303S05S}.
As same as discussed above,
the solar wind decreases the amount of the overlying magnetic flux 
and help the flux rope to escape from the Sun.

\subsection{Rotation of ejecting flux rope}
\label{dis2}

As described in \S \ref{res},
the rotations of ejecting flux ropes in 
the early  phase  in any cases
are caused by the writhe motion of 
the spheromak structure in the initial conditions.
However,  
the final angles of the escaping flux ropes 
are quite different ($\sim 90^\circ$) between in case A and C 
(Figure \ref{albta}a and \ref{albta}e).
The difference is caused by only 
the existence of the ambient field, 
i.e., the interaction between the ejecting flux rope and 
the ambient field as mentioned in \S \ref{para}.
The interaction is  
the reconnection of twisted field within the flux rope to
the untwisted ambient field. 
As a result of reconnection of not completely anti-parallel fluxes, 
strongly bent magnetic flux can be formed.
As the result of a relaxation process of the flux, 
the flux rope can be rotated.

\begin{figure}
\plotone{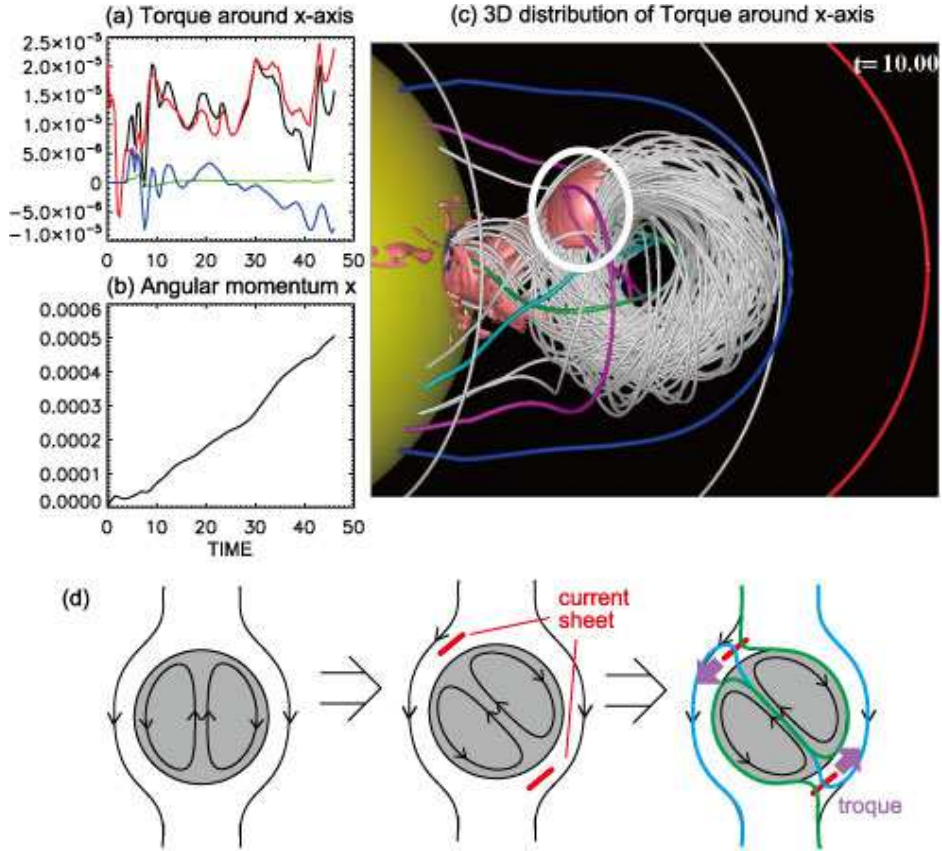}
\caption{Time variation of torques (a) 
and angular momentum (b) around $x$-axis in case C (parallel).
Green, blue, red, and black lines in panel (a)  
show time variations of torques 
caused by dynamic pressure, 
gas pressure gradient, Lorentz force, 
and total force, respectively. 
(c) Three dimensional distribution 
of the magnetic torque around the $x$ axis. 
Pink surfaces show positions 
where counter-clockwise (negative $x$) torques work.
White circle indicates the position of 
the net counter-clockwise torque which 
substantially rotates the flux rope.
(d) Schematic pictures of magnetic field lines
on a horizontal cross section in the ejecting structure.
Due to its writhe motion,
the twisted flux rope rotates counter-clockwise
(evolution form the left panel to the center panel).
As a result, 
current sheets (indicated with red lines in the center panel) 
are formed on the partial surface of the flux rope.
Reconnection in the sheets produce 
steep bend magnetic field 
(green and light blue lines in the right panel of (d) and
purple tube in panel (c)),
which causes the torque due to the magnetic tension force.
}
\label{torq}
\end{figure}

Angular momentum ($L_x$) and torque($\tau_x$) 
around the $x$ axis are obtained from
their integral in the space 
$V_1 \equiv  \left \{{\bf r} |\right. 
r_{\rm min} \leq r \leq   r_{\rm max}, 
 \pi/4 \leq \theta \leq 3 \pi /4, 
-\pi/4 \leq \phi \leq \left. \pi/4 
  \right\} $
\begin{equation}
    L_x = \int \int \int_{V_1} {\rm d} V
              \left( {\bf r} \times \rho {\bf v} \right)
              \cdot \hat{\bf x} 
\end{equation}
\begin{equation}
  \tau_x = \int \int \int_{V_1} {\rm d} V
              \left( {\bf r} \times {\bf F} \right)
              \cdot \hat{\bf x}  
\end{equation}
where ${\bf F}$ expresses any force
of Lorentz force, or gradients of gas pressure, or dynamic pressure.
The time variations of the torques around the $x$ axis 
are shown in Figure \ref{torq}a 
and that of angular momentum $x$ component is 
shown in Figure \ref{torq}b. 
Figures \ref{torq}a and \ref{torq}b show 
that the structure 
continues to be rotated  
dominantly by the torque due to the Lorentz force.
This result appears that the rotation caused by 
the writhe motion of the initial flux rope. 
However, Figure \ref{torq}c shows the 
three dimensional distribution of the torque. 
The net torque mainly works 
at the surface of the flux rope 
(indicated pink surfaces emphasized with a white circle 
in Figure \ref{torq}c). 
It appears that the rotation are caused by interaction 
between magnetic field inside the flux rope 
and the ambient magnetic field. 
Magnetic field line which penetrates the strong torque region
is drawn with ``purple" line in Figure \ref{torq}c.
The ``purple" line appears strongly wind,
i.e., just after reconnected.
This result means that 
the strong torque regions correspond to 
the current sheets where the reconnection 
between the flux rope and the ambient field occurs.

Figure \ref{torq}d shows a schematic picture of 
the evolution of the magnetic field lines in case C
mentioned above and in \S \ref{para}. % and \S \ref{evoaf}.
The expanding flux rope initially rotates counter-clockwise 
due to its writhe motion. 
As a result,
current sheets are formed at two point on the surface, 
where the anti-parallel field are contact
(indicated with red lines in Figure \ref{torq}d). 
Reconnection occurs in the current sheets
and then strongly wind field lines
(indicated with green and light blue lines in Figure \ref{torq}d)
 are formed.
The untwisting motion of the winding field  
causes a strong torque (purple arrows in Figure \ref{torq}d) 
which rotates the flux rope. 

In a few decades ago,
a magnetic field configuration 
of spheromak in equilibrium state under uniform field
has been applied to a laboratory plasma experiment,
where plasma must be confined within a device.
However, such a spheromak type field configuration 
is unstable to tilting mode perturbation
\citep{1983PhRvL..50...38S}.
The field configuration settles down where magnetic moment 
of the spheromak is parallel to the background field. 
In the nonlinear stage of the instability,
the magnetic field causes
reconnection which is the same as explained in Figure \ref{torq}d.
The numerical results in this study imply that 
such an effect in laboratory plasma 
works also in the coronal plasma.

Such rotations of  ejecting flux ropes are reported in 
some CME observations. 
\citet{2008ApJ...675L..49Y} investigated 
the axis angles of magnetic clouds, 
halo CME axes derived from coronagraph images, and 
EUV post-eruptive arcade axes in CME source regions.
They found that 
there are good correlation between the axes of EUV arcades
and CME angles within a deviation of $\pm 45^\circ$. 
The fact means that 
some fraction of ejected magnetic structures
are significantly rotated 
during the travels in the lower solar corona
but the amount of the rotation is not so much.  

The numerical results of the present work 
are consistent with the observation in that 
flux ropes can be rotated significantly 
by the interaction with closed field. 
However, the rotation angle in case C is more  
than that of most CMEs in \citet{2008ApJ...675L..49Y}.
The reason is that the effect of the ambient field in case C 
is much stronger than the real solar corona 
because of the following reasons:
The dipole field assumed   in this study
decreases in $r^{-3}$ with the distance from the Sun $r$, 
and is the potential field whose influence can reach most far away. 
Most of the dipole field flux are closed 
because solar wind is not taken into account.
Alfv{\'e}n speed in the simulation is very small  
because of the assumption of the high plasma density.
The slow evolutions allow 
that the ejecting structure can show enough rotation.
The condition in case C is an ideal case to examine 
the interaction between the ejecting flux rope 
and the ambient field.
Furthermore, 
although the time evolution in case C is very slow 
(the terminal CME speed $ \sim 40$km s$^{-1}$), 
the terminal speed is comparable to Alfv{\'e}n speed inside the flux rope. 
%The numerical results of the present work is not inappropriate.
%The slow evolutions allow that the ejecting structure can show 
%enough rotation in either case. 

The  rotation process of ejecting flux ropes is 
also numerically studied by 
\cite{2008JGRA..11309103G} and \cite{2009ApJ...697.1918L}. 
\cite{2008JGRA..11309103G} discussed 
the rotation process with
the numerical result of 
eruption of a twisted flux rope 
within a potential arcade.  
They showed that 
the rotation of the flux rope axis ($\sim 115^\circ$)
is caused by the writhe motion.  
\cite{2009ApJ...697.1918L} discussed 
the rotation precess with the results of
two eruptions of arcades sheared oppositely each other.   
They showed that the opposite chirality of the sheared arcades
leads the same amount of opposite rotations of 
the resulting flux ropes. 
In contrast to these studies, 
the results of the present study show that 
the rotation of the flux rope is affected by 
not only the writhe motion
but the interaction with ambient field 
as described in \S \ref{res}.
Contrast to their works, 
the results of the present study suggest that 
an ejecting flux rope can be
significantly rotated as a result  of reconnection  
with its ambient field.

\subsection{Evolution of Anchored Axial flux}
\label{evoaf}
In case C,   
because of the interchange reconnection 
mentioned in \S \ref{para}
the foot points of the ejecting flux rope 
appear to move outside 
(colored lines in Figure \ref{evoC} and \ref{angle}). 
In the initial condition 
only the ``green" line was included 
in the anchored axial flux of the ejecting flux rope,
but finally all the ``light blue", 
``blue" and ``red" lines are included.
These changes of anchored foot points   
are caused by a topological change of magnetic flux 
due to reconnection in two regions; 
at the surface and behind of the ejecting flux rope. 
As described in \S \ref{res},  
the ejecting and expanding flux rope forms 
$\Omega$-type loops of the initial axial flux. 
The first reconnection occurs in the current sheet 
at the center of the waist of the loop 
(see the ``green" line in Figure \ref{evoC} and \ref{angle}).
This reconnection severs the anchored axial flux and 
makes the ejecting flux rope structure  detached.

At the same time,  
the ejecting flux rope reconnects 
with the ambient field in its side 
(``light blue" and ``blue" lines) 
and front (the ``red" line).
In the initial condition of this case, 
the poloidal field of the flux rope 
is parallel and therefore unfavorable to reconnection.  
However the flux rope rotates 
due to its untwisting (writhe) motion,
as described in \S \ref{dis2}.
The reconnected ambient field is connected to
the inside of the flux rope, i.e.,
becomes the new axial flux of the ejecting flux rope.
The interchange of the anchored axial field 
makes distance between the legs larger 
and makes the foot point field strength weaker. 
As a result, the magnetic tension force can decrease and then
the flux rope becomes easier to escape from the Sun.

The interchange of anchored flux
explains the evolution of dimming region 
observed in many flares and/or CMEs
with soft X-ray or EUV.   
In many flares, strong dimming are often observed 
in both sides of the post eruptive arcades, 
and there are following weak dimming 
far from the flare cite
\citep{1997ApJ...491L..55S}.
Such strong dimming can be formed by 
a sudden expansion of the erupting core and 
upward dragged overlying flux
coupled with coincident rarefaction due to reconnection
\citep{2005ApJ...634..663S}. 
The following weak dimming
are formed by some interaction (stretched or reconnected) 
with the erupting structure. 
This evolution of the magnetic field structure is
similar to that mentioned by \cite{2008JGRA..11309103G}. 
The results of  case C in the present study 
also show that 
the weak dimming are seen in new footpoints  of axial flux formed   
due to reconnection described in this section. 

 \section{Summary}

In this study, we performed
three global MHD simulations of 
interaction (magnetic reconnection) between an ejecting flux rope 
and its ambient field, as  
a launch process of coronal mass ejections (CMEs).
The initial conditions are assumed as  
ejecting flux ropes
under the three different ambient fields:
A case without ambient field  
and cases with dipole field of two opposite directions 
which are parallel and anti-parallel to that of 
the flux rope surface. 
The flux rope can escape to 
the outer boundary of the numerical domain
(evolve to CMEs)
in the no ambient case and one dipole case 
where the direction of ambient field is parallel, 
while it cannot escape in the other dipole case.   

The flux ropes show rotations 
perpendicular to the ejecting direction, 
that are caused by not only untwisting (writhe) motion
of the flux rope but also the interaction between
the ejecting flux rope and the ambient field.
Especially, the results show 
the possibility of rotation due to
the relaxation of a complicated field structure
which results from the reconnection between 
the flux rope and the ambient field.
The reconnection and resulting rotation are similar to 
that seen in a spheromak experiment of laboratory plasma.

The results of this study show
that the interaction between 
an ejecting flux rope and its ambient field
is a significant process for determining 
CME formation and CME orientation.
For the achievement of space weather forecast
it is necessary to predict occurrence of a CME
when an flare occurs.
A powerful tool for such prediction is  
MHD simulation of the flare and 
accompanying ejecting flux rope
under a realistic coronal magnetic field.

\acknowledgments
DS thanks S. Inoue, R. Kataoka, and K. Hayashi 
for fruitful discussion. 
The numerical calculation was carried out using the Earth Simulator
(ES) in Japan Agency
for Marine-Earth Science and Technology (JAMSTEC), under the ES project
``Space and Earth System Modeling''  (PI: K. Kusano).
This work was supported by 
the Grant-in-Aid for Creative Scientific Research 
``The Basic Study of Space Weather Prediction" 
(17GS0208, Head Investigator: K. Shibata) 
from the Ministry of Education, Culture, Sports, Science, and
Technology (MEXT) of Japan
and in part
by the Grand-in-Aid for the Global COE program ``The Next
Generation of Physics, Spun from Universality and Emergence"
from MEXT of Japan.

\bibliography{shiota.bib}
%%\begin{thebibliography}{}
%%\end{thebibliography}

\end{document}